\documentclass[prb,twocolumn,superscriptaddress,showpacs,amsmath,amssymb]{revtex4}
\usepackage{amsfonts}
\usepackage{bbm}
\usepackage{graphicx}
\usepackage{dcolumn}
\usepackage{bm}

\begin{document}
\title{Double Andreev reflections and double normal reflections
in nodal-line semimetal-superconductor junctions}

\author{Qiang Cheng}
\affiliation{School of Science, Qingdao University of Technology, Qingdao, Shandong 266520, China}
\affiliation{International Center for Quantum Materials, School of Physics, Peking University, Beijing 100871, China}

\author{Zhe Hou}
\affiliation{International Center for Quantum Materials, School of Physics, Peking University, Beijing 100871, China}

\author{Qing-Feng Sun}
\email[]{sunqf@pku.edu.cn}
\affiliation{International Center for Quantum Materials, School of Physics, Peking University, Beijing 100871, China}
\affiliation{Collaborative Innovation Center of Quantum Matter, Beijing 100871, China}
\affiliation{CAS Center for Excellence in Topological Quantum Computation, University of Chinese Academy of Sciences, Beijing 100190, China}

\begin{abstract}
We study systematically the scattering processes and the conductance spectra
in nodal-line semimetal-superconductor junctions using the extended Blonder-Tinkham-Klapwijk theory.
The coexistence of peculiar quadruple reflections are found,
which are the specular normal reflection, the retro-normal reflection,
the specular Andreev reflection and the retro-Andreev reflection.
The incident angle dependence and the quasiparticle energy dependence
of the double normal reflections and the double Andreev reflections
are investigated under various of values of parameters
such as the interfacial barrier height, the chemical potentials and the orbital coupling strength.
It is found that the appearance and the disappearance of the reflections and their magnitudes can be controlled through tuning these parameters. The scattering mechanism for the reflections are analyzed in details from the viewpoint of the band structure. We also investigate the conductance spectra for the junctions, which show distinctive features and strong anisotropy about the orientation relationships of the nodal-line and interface. The unique scattering processes and conductance spectra found in the junctions are helpful in designing new superconducting electronic devices and searching for the nodal-line in new materials experimentally.

\end{abstract}
\maketitle

\section{\label{sec1}Introduction}
As three-dimensional gapless materials with point nodes, Dirac and Weyl semimetals have attracted tremendous research interest due to their nontrivially topological properties, connections with particle physics and novel responses to electric and magnetic fields \cite{Armitage}. In contrast, nodal-line semimetals (NLSMs) are another new type of three-dimensional topological material with band touching along lines in the momentum space \cite{Burkov, Chiu, Fang, Wang}. It is important that the nodal-lines in NLSM can evolve into Dirac nodes or Weyl nodes when the protecting symmetry is broken \cite{Weng, Fang2}. Recently, NLSMs of a ring-shaped nodal-line and the drumhead surface states become a hot topic in condensed matter physics due to the emerging new physical properties \cite{Yan, Araujo, Huh, Gao, Moors, Ma, Ruiz, Yang, Oroszlany, Liu, Behrends}. For example, the quantum oscillations of NLSMs exhibit the phase shift different from Weyl fermions \cite{Li}. The orbital susceptibility of NLSMs shows a stronger $\delta$-function singularity than that of Dirac and Weyl semimetals with point nodes \cite{Koshino}.

The distinctive features of NLSMs with a ring-shaped nodal-line also show themselves in the transport aspects \cite{Molina, Chen, Chen2, Barati, Hao}.
The transport in the ballistic regime are usually responsible for or can be used to probe the extraordinary phenomena \cite{Lv}. The anomalous transverse current in NLSM with small inversion breaking can be induced by an electric field,
which may be detected by the dumbbell device with a ballistic constriction \cite{Khokhlov}. The Klein tunneling in the single-particle ballistic scattering of NLSMs can be realized when the incident angle differing from $90^{\circ}$ and the transport properties show strong anisotropy related to the orientation of the crystallographic axis \cite{Rui}.
The unexpected nonuniversal conductance fluctuation is found in NLSM,
which amplitude rises as the increase of spin-orbit coupling strength \cite{Hu}.

Nonetheless, researches on scattering processes and conductance in NLSM in contact with a superconductor (SC) are still blank, which may provide specific signatures for the ring-shaped nodal-line. It is well known that the so-called retro-Andreev reflection (RAR) occurs in the ordinary normal metal-SC junctions \cite{Andreev}. The Andreev reflection dominates
the conductance of the junction when the bias is less than
the superconducting gap.\cite{addsun1,addsun2}
After the discovery of the two-dimensional gapless material, the specular Andreev reflection (SAR) is possible in the graphene-SC junctions, which dominates the subgap conductance under the weakly doped situation \cite{Beenakker,addref1,addsun3,addsun4,addref2}. Recently, it is proposed that the type-$\text{\uppercase\expandafter{\romannumeral2}}$ Weyl semimetal-SC junctions can host double Andreev reflections, both RAR and SAR, due to the band tilt. The evolution of the double Andreev reflections are also studied when the orientation between interface and the band tilt is changed \cite{Hou}. It should be noted that the normal reflection in the above junctions is always the specular one.

In this paper, we study the transport properties in the NLSM-SC junction with the ring-shaped nodal-line. We find the quadruple reflections can happen simultaneously, which are RAR, SAR, the specular normal reflection (SNR) and the retro-normal reflection (RNR). The probabilities of the double Andreev reflections and the double normal reflections strongly depend on the incident angle and energy of quasiparticles, the interfacial barrier height, the chemical potentials, the orbital coupling strength and the mutual orientation of interface and the nodal-line. The reflection types and their magnitudes can be regulated by changing one of the parameters. The scattering mechanism are analyzed in details from the point of the band structure. The conductance spectra for the nodal-line parallel to and perpendicular to interface exhibit strong anisotropy although they both have the zeros when the quasiparticle energy equals to the chemical potential in NLSM.

The rest of paper is organized as follows. In Sec.\ref{sec2}, we demonstrate the model and numerical results for the junction with the nodal-line perpendicular to interface. The Hamiltonians for NLSM and SC and wave functions in the two regions are given. The reflection properties are defined and the expression of conductance is derived. Their dependences on parameters are discussed detailedly. In Sec.\ref{sec3}, we present the formalism and results for the junction with nodal-line parallel to interface. Sec.\ref{sec4} concludes this paper.

\section{\label{sec2}Nodal-line perpendicular to interface}
\subsection{\label{subsec2.1}Model and formalism}
\begin{figure}[!htb]
\includegraphics[width=1.0\columnwidth]{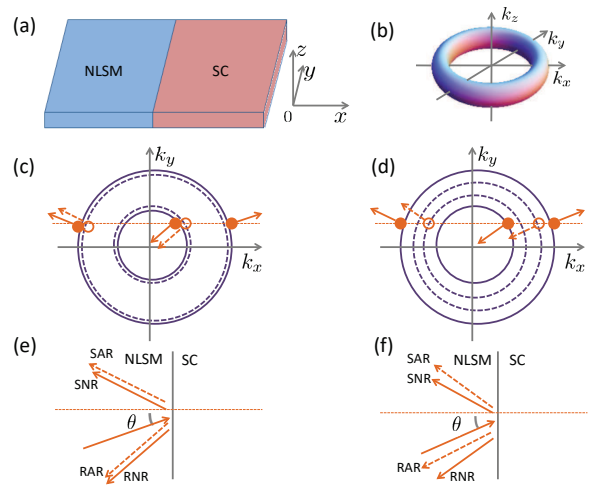}
\caption{
(a) Schematic illustration of the NLSM-SC junction with the nodal-line perpendicular to interface.
(b) The torus-like isoenergetic surface in NLSM.
Two concentric isoenergy-circles will be obtained if the surface
is cut by a plane parallel to the $k_x$-$k_y$ plane and
with a small value of $k_{z}$.
(c) The projection of the scattering processes on the $k_{x}$-$k_{y}$ plane.
The solid arrows denote the group velocity of ELQs (filled circles) and the dashed ones denote the group velocity of HLQs (empty circles). The isoenergetic circles for ELQs (solid lines) and HLQs (dashed lines) coincide with each other due to $\mu_{N}=0$. When $\mu_{N}\ne0$ and $\mu_{N}<\vert E_{z}\vert$, the isoenergetic circles no longer coincide but the scattering possesses the similar processes. (d) The projection of the scattering processes on the $k_{x}$-$k_{y}$ plane for $\mu_{N}>\vert E_{z}\vert$. The isoenergetic circles for ELQs and HLQs do not coincide. The retro-Andreev reflected HLQs occupy the big isoenergetic circle while the specular Andreev reflected HLQs occupy the small circle, which are different from the situation with $\mu_{N}<\vert E_{z}\vert$ in (c). (e) The projection of the scattering processes on the $x$-$y$ plane in the real space for $\mu_{N}=0$.
The gray solid line denotes the NLSM-SC interface and the symbol $\theta$ denotes the incident angle of ELQs. The arrows are parallel to those in (c). (f) The projection of the scattering processes on the $x$-$y$ plane in the real space for $\mu_{N}>\vert E_{z}\vert$. The arrows are parallel to those in (d).}
\label{fig1}
\end{figure}

We consider the NLSM-SC junctions as shown in Fig.\ref{fig1}(a),
which consist of the semi-infinite NLSM and SC.
The interface is located at $x=0$, which is parallel to the $y$-$z$ plane.
The interfacial barrier is modeled by a delta function $V(x)=V\delta(x)$.
The transport along the $x$ direction is considered.

We consider the two-orbit effective Hamiltonian for NLSM,
which is given by \cite{Xie,Chan}
\begin{equation}
\hat{H}_{N}(\bm{k})=\epsilon_{k}\hat{\sigma}_{z}-\hbar vk_{z}\hat{\sigma}_{y}-\mu_{N}\hat{\sigma}_{0},\label{HN}
\end{equation}
with $\epsilon_{k}=\frac{\hbar^2}{2m}(k_{x}^2+k_{y}^2+k_{z}^2)-E_{0}$. Here, the identity matrix $\hat{\sigma}_{0}$ and the Pauli matrices $\hat{\sigma}_{y}$ and $\hat{\sigma}_{z}$ are defined in the orbit space. The three component wave vector is $\bm{k}=(k_{x},k_{y},k_{z})$ and $\mu_{N}$ is the chemical potential. For $\mu_{N}=0$, the Hamiltonian describes the NLSM with a ring-shaped nodal-line in the plane $k_{z}=0$, which radius is given by $k=\sqrt{{2mE_{0}}/{\hbar^2}}$. For $\mu_{N}\neq 0$, the nodal-line will evolve into a torus-shaped Fermi surface as shown in Fig.\ref{fig1}(b).
Since the nodal-line is in the plane $k_{z}=0$, the transport properties in the $z$ direction is different from that
in the $x$ and $y$ directions, i.e. the NLSM with the Hamiltonian in Eq.(1)
has the strong anisotropy. Here we consider the transport along the $x$ direction
with the nodal-line perpendicular to the NLSM-SC interface at $x=0$ as shown in Fig.\ref{fig1}(a).
The transport along the $z$ direction with the nodal-line parallel to the NLSM-SC interface is studied in Sec.\ref{sec3}.

The Bogoliubov$-$de Gennes (BdG) Hamiltonian for NLSM can be written as
\begin{equation}
\check{H}_{N}(\bm{k})=\left(
\begin{array}{cc}
\hat{H}_{N}(\bm{k})&0\\
0&-\hat{H}^{*}_{N}(-\bm{k})
\end{array}\right),\label{BdGHN}
\end{equation}
in the orbit$\otimes$particle-hole space. Through solving the BdG equation $\check{H}_{N}(-i\nabla)\psi_{N}=E\psi_{N}$ with the substitution of $-i\nabla$ for $\bm{k}$ in $\check{H}_{N}(\bm{k})$, the energy dispersions for the electron-like quasiparticles (ELQs) and the hole-like quasiparticles (HLQs) can be obtained, which are
\begin{equation}
E_{e}^{\pm}=\pm\sqrt{\epsilon_{k}^2+E_{z}^2}-\mu_{N},
\end{equation}
and
\begin{equation}
E_{h}^{\pm}=\pm\sqrt{\epsilon_{k}^2+E_{z}^2}+\mu_{N},
\end{equation}
with $E_{z}=\hbar vk_{z}$ characterizing the orbital coupling strength.
From the dispersions, it is found that the gap $2\vert E_{z}\vert$ situated at the nodal-line is opened in the quasiparticle spectrum.
The gap spans from $-\vert E_{z}\vert-\mu_{N}$ to $\vert E_{z}\vert-\mu_{N}$ for ELQs.
For HLQs, the gap spans from $-\vert E_{z}\vert+\mu_{N}$ to $\vert E_{z}\vert+\mu_{N}$.

For $\mu_{N}=0$, the two gaps are both symmetric about the zero energy,
which span from $-\vert E_{z}\vert$ to $\vert E_{z}\vert$.
When the energy $E$ of the incident ELQs
is larger than $\vert E_{z}\vert$,
the conduction band $E_{e}^{+}$ and the valence band $E_{h}^{+}$ participate
in the scattering processes.
The isoenergetic circles with $E>\vert E_{z}\vert$
in the $k_{x}$-$k_{y}$ plane are shown in Fig.\ref{fig1}(c).
The isoenergy-circles for ELQs and for HLQs coincide in this situation. Considering that the incident energy $E$ and the wave vectors $k_y$ and $k_z$ (i.e. $E_z$)
are conserved in the scattering processes,
there are two types of ELQs (HLQs): one occupies the big circle with $\bm{k}=(k_{x}^{e(h)+},k_{y},k_{z})$ and the other occupies the small circle
with $\bm{k}=(k_{x}^{e(h)-},k_{y},k_{z})$, in which $k_{x}^{e,h\pm}=\sqrt{\frac{2m}{\hbar^2}(E_{0}\pm\sqrt{E^2-\hbar^2v^2k_{z}^{2}})-k_{y}^{2}-k_{z}^{2}}$.

If we denote the group velocity as $\bm{v}=(v_{x},v_{y},v_{z})$, it can be found that $(v_{x},v_{y},v_{z})$ have the same sign with $(k_{x}^{e(h)+},k_{y},k_{z})$ and $(-k_{x}^{e(h)-},-k_{y},k_{z})$.
For the injection of ELQs from NLSM, there are two types of the incident ELQs
with the wave vectors being $\bm{k}=(k_{x}^{e+},k_{y},k_{z})$ and
$\bm{k}=(-k_{x}^{e-},k_{y},k_{z})$.
However, when a beam of ELQs are injected from NLSM, regardless of their
wave vector being $k_{x}^{e+}$ or $-k_{x}^{e-}$,
there will always be four reflection processes (see Figs.\ref{fig1}(c) and (e)):
double Andreev reflections (SAR and RAR)
with the wave vector $\bm{k}=(-k_{x}^{h+},k_{y},k_{z})$ and
$(k_{x}^{h-},k_{y},k_{z})$
and double normal reflections (SNR and RNR)
with $\bm{k}=(-k_{x}^{e+},k_{y},k_{z})$ and
$(k_{x}^{e-},k_{y},k_{z})$.

Taking the injection of an ELQ with energy $E$ and the wave vector $k_{x}^{e+}$ as an example, the resulting wave function in NLSM can be solved from the BdG equation and written as
\begin{equation}
\begin{split}
\psi_{N}(x<0)=&\left(
\begin{array}{c}
i\chi_{1}\\
1\\
0\\
0
\end{array}\right)e^{ik_{x}^{e+}x}
+
r_{n1}\left(
\begin{array}{c}
i\chi_{1}\\
1\\
0\\
0
\end{array}\right)e^{-ik_{x}^{e+}x}\\
+
r_{n2}&\left(
\begin{array}{c}
i\chi_{2}\\
1\\
0\\
0
\end{array}\right)e^{ik_{x}^{e-}x}
+
r_{a1}\left(
\begin{array}{c}
0\\
0\\
-i\chi_{2}\\
1
\end{array}\right)e^{-ik_{x}^{h+}x}\\
+r_{a2}&\left(
\begin{array}{c}
0\\
0\\
-i\chi_{1}\\
1
\end{array}\right)e^{ik_{x}^{h-}x},
\end{split}\label{wf1}
\end{equation}
with $\chi_{1}=(E+\Omega_{N})/E_{z}$, $\chi_{2}=(E-\Omega_{N})/E_{z}$ and $\Omega_{N}=\sqrt{E^2-E_{z}^2}$. The symbols $r_{n1},r_{n2},r_{a1}$ and $r_{a2}$ represent reflection amplitudes for SNR, RNR, SAR and RAR, respectively.

For $\mu_{N}\neq0$, neither of the two gaps is symmetric about $E=0$. We discuss this situation in two aspects. The first aspect is $\mu_{N}<\vert E_{z}\vert$. In this case, the gap for ELQs spans from $-\vert E_{z}\vert-\mu_{N}<0$ to $\vert E_{z}\vert-\mu_{N}>0$, while that for HLQs spans from $-\vert E_{z}\vert+\mu_{N}<0$ to $\vert E_{z}\vert+\mu_{N}>0$. The isoenergetic circles for ELQs and HLQs split and the big circle for ELQs (HLQs) becomes bigger (smaller) and the small circle for ELQs (HLQs) becomes smaller (bigger). When $E>\vert E_{z}\vert+\mu_{N}$, the bands involved in the scattering processes are still $E_{e}^{+}$ and $E_{h}^{+}$ and the reflections are similar to those given in Figs.\ref{fig1}(c) and (e). The specular Andreev reflected HLQs still occupy the big circle and the retro-Andreev reflected HLQs still occupy the small circle.

Consider also the injection of an ELQ with energy $E$ and  the wave vector $k_{x}^{e+}$, the resulting wave function in NLSM is
\begin{equation}
\begin{split}
\psi_{N}(x<0)=&\left(
\begin{array}{c}
i\chi_{11}\\
1\\
0\\
0
\end{array}\right)e^{ik_{x}^{e+}x}
+
r_{n1}\left(
\begin{array}{c}
i\chi_{11}\\
1\\
0\\
0
\end{array}\right)e^{-ik_{x}^{e+}x}\\
+
r_{n2}&\left(
\begin{array}{c}
i\chi_{12}\\
1\\
0\\
0
\end{array}\right)e^{ik_{x}^{e-}x}
+
r_{a1}\left(
\begin{array}{c}
0\\
0\\
-i\chi_{22}\\
1
\end{array}\right)e^{-ik_{x}^{h+}x}\\
+r_{a2}&\left(
\begin{array}{c}
0\\
0\\
-i\chi_{21}\\
1
\end{array}\right)e^{ik_{x}^{h-}x},
\end{split}\label{wf2}
\end{equation}
with $\chi_{11(12)}=(E+\mu_{N}+(-)\Omega_{N}^{+})/E_{z}$, $\chi_{21(22)}=(E-\mu_{N}+(-)\Omega_{N}^{-})/E_{z}$, $k_{x}^{e\pm}=\sqrt{\frac{2m}{\hbar^2}(E_{0}\pm\Omega_{N}^{+})-k_{y}^2-k_{z}^2}$, $k_{x}^{h\pm}=\sqrt{\frac{2m}{\hbar^2}(E_{0}\pm\Omega_{N}^{-})-k_{y}^2-k_{z}^2}$ and $\Omega_{N}^{\pm}=\sqrt{(E\pm\mu_{N})^2-E_{z}^2}$. The coefficients
$r_{n1}$, $r_{n2}$, $r_{a1}$ and $r_{a2}$
have the same meanings as those in Eq.(\ref{wf1}). As $\mu_{N}\rightarrow0$, the wave function in Eq.(\ref{wf2}) degenerates into that in Eq.(\ref{wf1}).

The second aspect for $\mu_{N}\neq0$ is $\mu_{N}>\vert E_{z}\vert$. The gap for ELQs spans from $-\vert E_{z}\vert-\mu_{N}<0$ to $\vert E_{z}\vert-\mu_{N}<0$ and that for HLQs spans from $\mu_{N}-\vert E_{z}\vert>0$ to $\mu_{N}+\vert E_{z}\vert>0$. Therefore, when $0<E<\mu_{N}-\vert E_{z}\vert$, the conduction bands $E_{e}^{+}$ and $E_{h}^{-}$ participate in the scattering processes. In this situation, the isoenergetic circles for ELQs and HLQs are shown in Fig.\ref{fig1}(d). The reflections possess a little difference from those with $\mu_{N}<\vert E_{z}\vert$. The incident ELQs will be specular Andreev reflected as HLQs with the wave vector $(-k_{x}^{h-},k_{y},k_{z})$ on the small circle and will be retro-Andreev reflected as HLQs with the wave vector $(k_{x}^{h+},k_{y},k_{z})$ on the big circle (see Figs.\ref{fig1}(d) and (f)). The wave function in NLSM is
\begin{equation}
\begin{split}
\psi_{N}(x<0)=&\left(
\begin{array}{c}
i\chi_{11}\\
1\\
0\\
0
\end{array}\right)e^{ik_{x}^{e+}x}
+
r_{n1}\left(
\begin{array}{c}
i\chi_{11}\\
1\\
0\\
0
\end{array}\right)e^{-ik_{x}^{e+}x}\\
+
r_{n2}&\left(
\begin{array}{c}
i\chi_{12}\\
1\\
0\\
0
\end{array}\right)e^{ik_{x}^{e-}x}
+
r_{a1}\left(
\begin{array}{c}
0\\
0\\
-i\chi_{21}\\
1
\end{array}\right)e^{-ik_{x}^{h-}x}\\
+r_{a2}&\left(
\begin{array}{c}
0\\
0\\
-i\chi_{22}\\
1
\end{array}\right)e^{ik_{x}^{h+}x},
\end{split}\label{wf3}
\end{equation}
where $\chi_{11}$, $\chi_{12}$, $\chi_{21}$, $\chi_{22}$, $k_{x}^{e\pm}$ and $k_{x}^{h\pm}$ are the same as those in Eq.(\ref{wf2}). Now, if one increases the energy of the incident ELQs to $E>\mu_{N}+\vert E_{z}\vert$, the involved bands in the scattering processes become again the conduction band $E_{e}^{+}$ and the valence band $E_{h}^{+}$. In this case, the wave function in NLSM is still given by Eq.(\ref{wf2}).

The superconductivity in NLSM can be induced by its high-quality contact with SC. In this paper, we consider the spin-singlet $s$-wave pairing realized in the superconducting NLSM. The BdG Hamiltonian for the SC formed in NLSM
at the region of $x>0$ is
\begin{equation}
\check{H}_{S}(\bm{k})=\left(
\begin{array}{cc}
\hat{H}_{S}(\bm{k})&\hat{\Delta}\\
\hat{\Delta}&-\hat{H}^{*}_{S}(-\bm{k})
\end{array}\right),\label{HS}
\end{equation}
in which $\hat{H}_{S}=\epsilon_{k}\hat{\sigma}_{z}-\hbar vk_{z}\hat{\sigma}_{y}-\mu_{S}\hat{\sigma}_{0}$ and $\hat{\Delta}=\Delta\hat{\sigma}_{0}$ with $\mu_{S}$ the chemical potential in SC and $\Delta$ the superconducting gap magnitude.

By solving the BdG equation for SC, the wave function in the superconducting region can be written as
\begin{equation}
\begin{split}
\psi_{S}(x>0)=&
t_{1}\left(
\begin{array}{c}
iu\eta_{11}\\
u\\
i\eta_{11}\\
1
\end{array}\right)e^{ip_{x}^{+}x}
+
t_{2}\left(
\begin{array}{c}
iu\eta_{12}\\
u\\
i\eta_{12}\\
1
\end{array}\right)e^{-ip_{x}^{-}x}\\
+&
t_{3}\left(
\begin{array}{c}
iv\eta_{21}\\
v\\
i\eta_{21}\\
1
\end{array}\right)e^{-iq_{x}^{+}x}
+t_{4}\left(
\begin{array}{c}
iv\eta_{22}\\
v\\
i\eta_{22}\\
1
\end{array}\right)e^{iq_{x}^{-}x},
\end{split}\label{wfs3}
\end{equation}
where $u=(E+\Omega_S)/\Delta$, $v=(E-\Omega_S)/\Delta$, $\eta_{11(12)}=(\mu_{S}+\Omega_S +(-)\Omega_{S}^{+})/E_{z}$, $\eta_{21(22)}=(\mu_{S}-\Omega_S +(-)\Omega_{S}^{-})/E_{z}$ with $\Omega_S=\sqrt{E^2-\Delta^2}$ and $\Omega_{S}^{\pm}=\sqrt{(\Omega_S\pm\mu_{S})^2-E_{z}^2}$. The wave vector $p_{x}^{\pm}$ and $q_{x}^{\pm}$ are given by $p_{x}^{\pm}=\sqrt{\frac{2m}{\hbar^2}(E_{0}\pm\Omega_{S}^{+})-k_{y}^2-k_{z}^2}$ and $q_{x}^{\pm}=\sqrt{\frac{2m}{\hbar^2}(E_{0}\pm\Omega_{S}^{-})-k_{y}^2-k_{z}^2}$. The symbols $t_{1}$, $t_{2}$, $t_{3}$ and $t_{4}$ denote the transmission amplitudes of quasiparticles.

The reflection and transmission amplitudes can be obtained using the boundary conditions at $x=0$, which are
\begin{equation}
\begin{split}
\psi_{N}(x=0^{-})&=\psi_{S}(x=0^{+}),\\
\psi^{'}_{S}(x=0^{+})&-\psi^{'}_{N}(x=0^{-})=\frac{2mV}{\hbar^2}\check{M}\psi_{N}(x=0),
\end{split}
\end{equation}
in which $\check{M}$ is a $4\times4$ diagonal matrix with the diagonal elements $(1,-1,1,-1)$.

The probabilities for SNR and RNR can be defined as
\begin{equation}
\begin{split}
R_{n1}(E)&=\vert r_{n1}\vert^2,\\
R_{n2}(E)&=\text{Re}\Big[\frac{k_{x}^{e-}}{k_{x}^{e+}}\Big]\vert r_{n2}\vert^2\Big\vert\frac{\chi_{12}^2-1}{\chi_{11}^2-1}\Big\vert.
\end{split}
\end{equation}
The probabilities for SAR and RAR can be defined as
\begin{equation}
\begin{split}
R_{a1}(E)&=\text{Re}\Big[\frac{k_{x}^{h+}}{k_{x}^{e+}}\Big]\vert r_{a1}\vert^2\Big\vert\frac{\chi_{22}^2-1}{\chi_{11}^2-1}\Big\vert,\\
R_{a2}(E)&=\text{Re}\Big[\frac{k_{x}^{h-}}{k_{x}^{e+}}\Big]\vert r_{a2}\vert^2\Big\vert\frac{\chi_{21}^2-1}{\chi_{11}^2-1} \Big\vert,
\end{split}
\end{equation}
for $\mu_{N}<\vert E_{z}\vert$ or $\mu_{N}>\vert E_{z}\vert$ and $E>\mu_{N}+\vert E_{z}\vert$, and
\begin{equation}
\begin{split}
R_{a1}(E)&=\text{Re}\Big[\frac{k_{x}^{h-}}{k_{x}^{e+}}\Big]\vert r_{a1}\vert^2\Big\vert\frac{\chi_{21}^2-1}{\chi_{11}^2-1}\Big\vert,\\
R_{a2}(E)&=\text{Re}\Big[\frac{k_{x}^{h+}}{k_{x}^{e+}}\Big]\vert r_{a2}\vert^2\Big\vert\frac{\chi_{22}^2-1}{\chi_{11}^2-1} \Big\vert,
\end{split}
\end{equation}
for $\mu_{N}>\vert E_{z}\vert$ and $E<\mu_{N}-\vert E_{z}\vert$.

The reflection amplitudes $\tilde{r}_{n1}$, $\tilde{r}_{n2}$, $\tilde{r}_{a1}$, $\tilde{r}_{a2}$ and the corresponding probabilities $\tilde{R}_{n1}$, $\tilde{R}_{n2}$, $\tilde{R}_{a1}$, $\tilde{R}_{a2}$
for the injection of an ELQ with the wave vector $-k_{x}^{e-}$
can also be calculated by using the same way.
The defined probabilities satisfy the conservation conditions,
\begin{equation}
\begin{split}
R_{n1}+R_{n2}+R_{a1}+R_{a2}=1,\\
\tilde{R}_{n1}+\tilde{R}_{n2}+\tilde{R}_{a1}+\tilde{R}_{a2}=1,\label{pc}
\end{split}
\end{equation}
inside the gap, i.e. $E<\Delta$, due to the absence of quasiparticle transmissions.

According to the Blonder$-$Tinkham$-$Klapwijk formalism \cite{Blonder}, the conductance $\sigma$ can be expressed as
\begin{equation}
\sigma=\frac{2e^2}{h}\frac{S}{(2\pi)^2}\frac{\sqrt{2mE_{0}}}{\hbar^2v}
{\int_{-\frac{\pi}{2}}^{\frac{\pi}{2}}\int_{-eV_b}^{eV_b}(\sigma_{1}+\sigma_{2})
\cos{\theta}d\theta dE_{z}},\label{cond}
\end{equation}
in which $\sigma_{1}(eV_b)=1+R_{a1}(eV_b)+R_{a2}(eV_b)-R_{n1}(eV_b)-R_{n2}(eV_b)$, $\sigma_{2}(eV_b)=1+\tilde{R}_{a1}(eV_b)+\tilde{R}_{a2}(eV_b)-\tilde{R}_{n1}(eV_b)
-\tilde{R}_{n2}(eV_b)$,
$V_b$ is the bias on the NLSM-SC junction,
$S$ is the cross-sectional area of the junction and $\theta$ is
the incident angle in the $x$-$y$ plane for ELQs (see Figs.\ref{fig1}(e) and (f)). The factor $2$ in $2e^2/h$ accounts for the spin degeneracy of energy bands in NLSM.
The reflection probabilities are even functions of $\theta$ and $E_{z}$, so do $\sigma_{1}$ and $\sigma_{2}$. The normalized conductance can be defined by $\sigma/\sigma_{0}$ with $\sigma_{0}$ the conductance of the NLSM-NLSM junction.

\subsection{\label{subsec2.2}Results and discussions}

We define the wave vector $k_{0}=\sqrt{2m\Delta/\hbar^2}$
and the effective barrier height can be given by $V_{0}=\frac{2mV}{\hbar^2k_{0}}$.
Throughout the calculations, the energy $E_{0}$ characterizing the size of the nodal-line
is taken as $200\Delta$.
We also take $\mu_{N}=\mu_{S}$ which will eliminate the wave vector mismatch
caused by the different chemical potentials in NLSM and SC.
If $\mu_{S}\not=\mu_{N}$ and $\mu_{S}$ much larger than $\Delta$,
only the probabilities of the reflections are slightly affected,
but double normal-reflections and double Andreev reflections are always present.
For clarity, only the numerical results for the scattering processes associated
with the incident ELQs having the wave vector $k_{x}^{e+}$ are presented and discussed.
The results for the incident ELQs with the wave vector $-k_{x}^{e-}$ can be obtained
and analyzed in a similar way.
Next, we will discuss the incident angle dependence and
the energy dependence of the reflection probabilities and the conductance spectra in details.

\subsubsection{\label{subsubsec2.2.1}The incident angle dependence of reflections}

\begin{figure}[!htb]
\includegraphics[width=1.0\columnwidth]{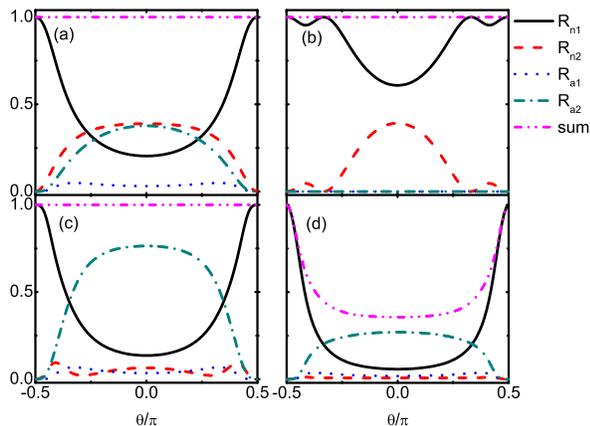}
\caption{The $\theta$ dependence of reflection probabilities
for $\mu_{N}=\mu_{S}=0.5\Delta$, $E_{z}=0.3\Delta$ and $V_{0}=10$
with (a) the incident energy $E=0.1$, (b) $E=0.5\Delta$, (c) $E=0.9\Delta$,
and (d) $E=1.2\Delta$.}
\label{fig2}
\end{figure}

First, we present the reflection probabilities as functions of
the incident angle $\theta$, which are always symmetric about $\theta=0$.
Fig.\ref{fig2} shows the results for different incident energy $E$
at $E_{z}=0.3\Delta$ and $\mu_{N}=\mu_{S}=0.5\Delta$.
In this instance, the band gap for ELQs spans from $-0.8\Delta$ to $-0.2\Delta$
and that for HLQs spans from $0.2\Delta$ to $0.8\Delta$.
When $0<E<0.2\Delta$, ELQs on the conduction band $E_{e}^{+}$
and HLQs on the conduction band $E_{h}^{-}$ participate
in the scattering processes which are schematically
presented in Figs.\ref{fig1}(d) and (f).
Double Andreev reflections and double normal reflections exist simultaneously.
The four reflection probabilities are all non-zero
as shown in Fig.\ref{fig2}(a) when $E=0.1\Delta$.
As $\theta$ increases from $0$ to $\pi/2$,
SNR $R_{n1}$ is enhanced while RNR $R_{n2}$ and SAR $R_{a1}$ are weakened.
As for RAR $R_{a2}$, it exhibits a small oscillation.
The curve ``sum" gives the summation of the reflection probabilities.
Its value is always $1$ for $E<\Delta$ according to
the conservation relations in Eq.(\ref{pc}).

When $0.2\Delta<E<0.8\Delta$, only ELQs on the conduction band $E_{e}^{+}$
take part in the scattering processes.
For the incident energy $E=0.5\Delta$ in Fig.\ref{fig2} (b),
the energy lying in the gap for HLQs leads to the vanishing SAR and RAR
and $R_{a1}=R_{a2}=0$ exactly.
Double normal reflections still exist as two inverse phase oscillating curves.
SNR $R_{n1}$ and RNR $R_{n2}$ reach their valley value
and peak value at $\theta=0$, respectively.
When $E>0.8\Delta$, HLQs on the valence band $E_{h}^{+}$ will be activated
and participate in the scattering processes
which are similar to Figs.\ref{fig1}(c) and (e).
Double Andreev reflections and double normal reflections
reappear as shown in Fig.\ref{fig2}(c).
In the four reflections, RAR $R_{a2}$ becomes the dominant process
in a large angle range around $\theta=0$.
If one continuously increases the value of $E$ until $E>\Delta$
as given in Fig.\ref{fig2} (d) with $E=1.2\Delta$,
the reflections will be weakened since quasiparticle transmissions begin to happen. The conservation relations in Eq.(\ref{pc}) should be revised.
In this case, the summation of the reflection and transmission
probabilities equals to $1$.
From the above results, it is found that the reflection styles
and magnitudes can be controlled by adjusting the incident quasiparticle energy.

\begin{figure}[!htb]
\includegraphics[width=1.0\columnwidth]{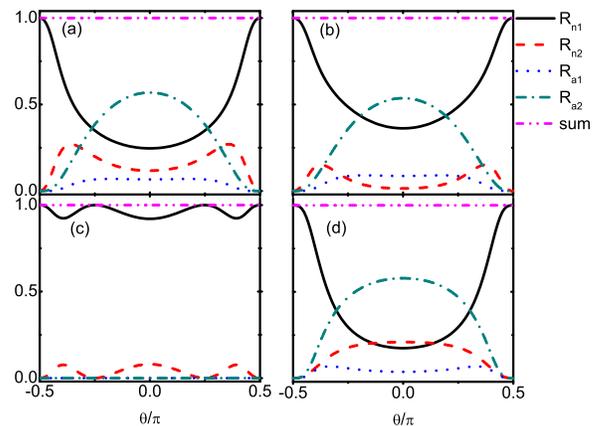}
\caption{The $\theta$ dependence of reflection probabilities
for $E=0.6\Delta$, $E_{z}=0.3\Delta$ and $V_{0}=10$ with
(a) $\mu_{N}=\mu_{S}=0$, (b) $\mu_{N}=\mu_{S}=0.2\Delta$, (c) $\mu_{N}=\mu_{S}=0.4\Delta$, and (d) $\mu_{N}=\mu_{S}=1.4\Delta$.}
\label{fig3}
\end{figure}

Fig.\ref{fig3} shows the reflection probabilities as functions of
the incident angle $\theta$ for different chemical potentials $\mu_{N/S}$.
The incident quasiparticle energy is taken as $E=0.6\Delta$ and
$E_{z}$ is fixed at $0.3\Delta$.
For $\mu_{N}=\mu_{S}=0$, ELQs and HLQs in NLSM have the same energy dispersions.
The isoenergetic circles for ELQs and HLQs in the $k_{x}$-$k_{y}$
plane coincide with each other as presented in Fig.\ref{fig1} (c).
The energy gap is symmetric about $E=0$.
The ELQs on the conduction band $E_{e}^{+}$
and HLQs on the valence band $E_{h}^{+}$ participate in
the scattering processes which are shown in Figs.\ref{fig1}(c) and (e).
When $E < E_{z}$, all reflections disappear
due to the absence of the incident states in the gap.
But, as long as $E>E_{z}$, there will be double Andreev reflections
and double normal reflections (see Fig.\ref{fig3} (a)).
The variations of SNR $R_{n1}$ and RAR $R_{a2}$ are complementary.
That is to say, SNR will acquire its maximum (minimum) value
if RAR gets its minimum (maximum) value.
Actually, this is an universal character when four types of reflections coexist.
RNR $R_{n2}$ exhibits oscillating behaviour and obtains its peak value
near $\theta=\pi/2$. In contrast, the magnitude of SAR $R_{a1}$
is higher near $\theta=0$.

For $\mu_{N}=\mu_{S} > 0$, the energy bands for ELQs and HLQs split.
The energy for ELQs is lowered and that for HLQs is raised.
The energy gap is asymmetric about $E=0$.
When $\mu_{N}=\mu_{S}=0.2\Delta$, the gap for ELQs spans
from $-0.5\Delta$ to $0.1\Delta$ and
that for HLQs spans from $-0.1\Delta$ to $0.5\Delta$.
The bands involved in the scattering processes are still
the conduction band $E_{e}^{+}$ and valence band $E_{h}^{+}$
since $\mu_{N}<E_{z}$.
The four reflections are presented in Fig.\ref{fig3}(b).
Comparing with Fig.\ref{fig3}(a), the changes of double Andreev reflections
are not obvious while SNR and RNR are slightly elevated and reduced, respectively.
The probability for SAR $R_{a1}$ is obviously larger than RNR $R_{n2}$
in a wide angle range around $\theta=0$.

For $\mu_{N}=\mu_{S}=0.4\Delta$, the gap for ELQs is from $-0.7\Delta$ to $-0.1\Delta$ and that for HLQs is from $0.1\Delta$ to $0.7\Delta$.
The incident energy $E=0.6\Delta$ is just in the gap of HLQs and
crosses the conduction band $E_{e}^{+}$ of ELQs.
Therefore, only ELQs will be reflected and the Andreev reflections
vanish with $R_{a1}=R_{a2}=0$ exactly.
The probabilities of the double normal reflections are given
in Fig.\ref{fig3}(c), which exhibit inverse phase oscillations.
When we increase the chemical potentials,
the lower edge of the HLQs gap moves up
while the upper edge of the ELQs gap moves down.
If $\mu_{N}=\mu_{S}=0.9\Delta$, the energy $E=0.6\Delta$ still
crosses the conduction band $E_{e}^{+}$ and
at the same time it is tangent to the lower edge of the HLQs gap.
For the lager values of the chemical potentials
such as $\mu_{N}=\mu_{S}=1.4\Delta$ in Fig.\ref{fig3}(d),
the energy $E=0.6\Delta$ will simultaneously intersect
the conduction bands $E_{e}^{+}$ and $E_{h}^{-}$
and double Andreev reflections reoccur.
In this case, both the incident ELQs and outgoing HLQs are in the
conduction bands.
Here RNR $R_{n2}$ reaches its peak value at $\theta=0$
and RAR $R_{a2}$ is suppressed slightly around $\theta=0$
comparing to Figs.\ref{fig3}(a) and (b).
From the above analyses, it is found that the positions of band gaps
and hence the reflection processes in NLSM-SC junctions
can be tuned by the chemical potential which can be controlled by
an gate voltage.

\begin{figure}[!htb]
\includegraphics[width=1.0\columnwidth]{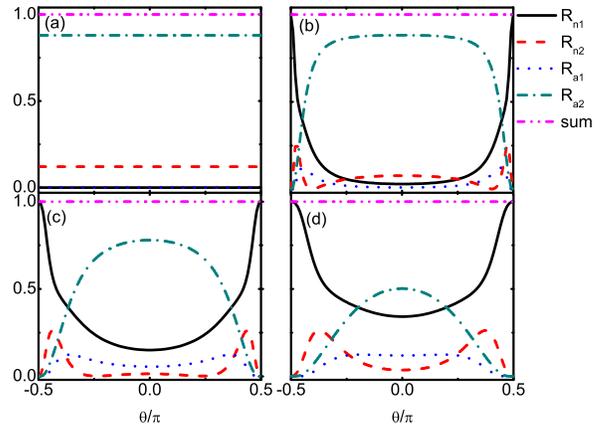}
\caption{The $\theta$ dependence of reflection probabilities
for $E=0.7\Delta$, $\mu_{N}=\mu_{S}=0.2\Delta$ and $E_{z}=0.4\Delta$
with (a) the barrier potential $V_{0}=0$, (b) $V_{0}=2$, (c) $V_{0}=5$, and (d) $V_{0}=10$.}
\label{fig4}
\end{figure}

Fig.\ref{fig4} gives the incident angle dependence of reflection probabilities
for different interfacial barrier heights.
We have taken $E_{z}=0.4\Delta$, $\mu_{N}=\mu_{S}=0.2\Delta$ and $E=0.7\Delta$.
Under these parameters, the conduction band $E_{e}^{+}$
and valence band $E_{h}^{+}$ both participate in the scattering.
However, for the transparent interface with $V_{0}=0$,
there are only retro-reflections (RNR and RAR) as shown in Fig.\ref{fig4}(a).
In particular, the SNR disappears also, which is similar
to the ordinary normal-metal-SC junction.\cite{Blonder}
The probabilities are approximatively independent of the incident angle.
In fact, the parameters $\mu_N$, $\mu_S$, the incident energy $E$,
and $E_z$ in Fig.4(a) are much smaller than $E_0=200\Delta$, leading to
the wave vectors $k_x^{e\pm}$ and $k_x^{h\pm}$ in NLSM
and $p_x^{\pm}$ and $q_x^{\pm}$ for quasiparticles in SC
can well be simplified into
$\sqrt{\frac{2m}{\hbar^2}E_{0}-k_{y}^2} = \frac{2m}{\hbar^2}E_{0} \cos{\theta}$.
Under this simplification and $V_{0}=0$,
the RNR and RAR coefficients can be analytically obtained as
\begin{eqnarray}
r_{n2}&=&\frac{v(\eta_{21}-\chi_{11})(\eta_{11}+\chi_{21})-u(\eta_{11}-\chi_{11})(\eta_{21}+\chi_{21})}
{-v(\eta_{21}-\chi_{12})(\eta_{11}+\chi_{21})+u(\eta_{11}-\chi_{12})(\eta_{21}+\chi_{21})},\\
r_{a2}&=&\frac{-(\eta_{11}-\eta_{21})(\chi_{11}-\chi_{12})}
  {-v(\eta_{21}-\chi_{12})(\eta_{11}+\chi_{21})+u(\eta_{11}-\chi_{12})(\eta_{21}+\chi_{21})},
\end{eqnarray}
which are independent of the incident angle $\theta$,
and SNR and SAR coefficients $r_{n1}=r_{a1}=0$ exactly.
As a fact, a finite barrier height is the necessary condition
for the formation of the specular reflections.
Even for a small barrier height with $V_{0}=2$ (see Fig.\ref{fig4}(b)),
the specular reflections will immediately appear
although they are not dominant in a large angle range centered at $\theta=0$.
When the barrier height is raised,
SNR $R_{n1}$ and SAR $R_{a1}$ are markedly enhanced and
the curve of SAR $R_{a1}$ exceeds that of RNR $R_{n2}$
as shown in Fig. \ref{fig4}(c) with $V_{0}=5$ in a large range.
When $V_{0}=10$, the probabilities of SNR and SAR at $\theta=0$
can achieve about $34\%$ and $12\%$, respectively.
The presence of finite barrier height is beneficial to
the specular reflections and detrimental to the retro-reflections.
The reflection styles and magnitudes can be regulated
by the interfacial transparency.

\begin{figure}[!htb]
\includegraphics[width=1.0\columnwidth]{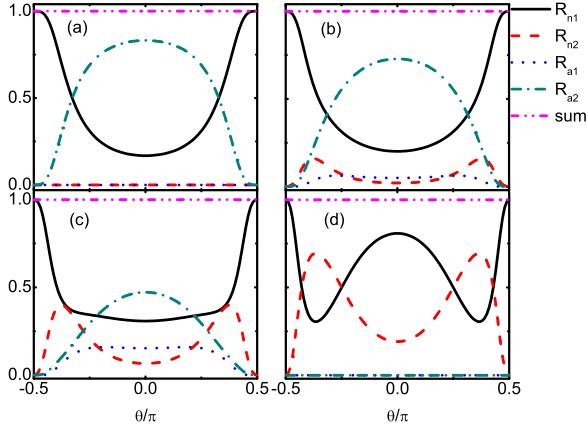}
\caption{
The $\theta$ dependence of reflection probabilities
for $E=0.8\Delta$, $\mu_{N}=\mu_{S}=0.2\Delta$ and $V_{0}=10$
with (a) $E_{z}=0$, (b) $E_{z}=0.3\Delta$, (c) $E_{z}=0.5\Delta$,
and (d) $E_{z}=0.7\Delta$.}
\label{fig5}
\end{figure}

Fig.\ref{fig5} gives the reflection probabilities
as functions of the incident angle $\theta$ for
different $E_{z}$ (i.e., the incident wave vector $k_z$).
For $E_{z}=0$, the orbital coupling in the Hamiltonian $\check{H}_{N}$
(see Eq.(\ref{HN})) and the gaps for ELQs and HLQs vanish.
The energy dispersions degenerate to $E_{e}^{\pm}=\pm\epsilon_{k}-\mu_{N}$
and $E_{h}^{\pm}=\pm\epsilon_{k}+\mu_{N}$.
When ELQs on the conduction band $E_{e}^{+}$ are injected from NLSM
to the NLSM-SC interface,
they will be normal reflected as ELQs on $E_{e}^{+}$
and Andreev reflected as HLQs on the valence $E_{h}^{-}$.
In other words, only two reflections, SNR and RAR,
can happen as shown in Fig.\ref{fig5}(a).
In this situation, the scattering processes in NLSM-SC junctions
are the same with those in the ordinary normal metal-SC junctions.\cite{Blonder}
For $E_{z}\ne0$, the coupling of two orbits
leads to the appearances of RNR and SAR.
Double normal reflections and double Andreev reflections
are realized as shown in Figs.\ref{fig5}(b) and (c).
The conduction band $E_{e}^{+}$ and the valence band $E_{h}^{+}$
are involved in the scattering.
With the increase of $E_{z}$, RNR $R_{n2}$ and SAR $R_{a1}$ are enhanced.
However for $E_{z}=0.7\Delta$ and $\mu_{N}=\mu_{S}=0.2\Delta$,
the situation is different.
Under these parameters, the energy $E=0.8\Delta$ lies in the gap of HLQs
and crosses the band $E_{e}^{+}$ of ELQs.
The Andreev reflections (RAR and SAR) disappear and
only the normal reflections of ELQs are possible.
The curves of SNR and RNR oscillate with the incident angle $\theta$,
which have two valley values and two peak values respectively
as shown in Fig.\ref{fig5}(d).

\subsubsection{\label{subsubsec2.2.2}The energy dependence of reflections}

\begin{figure}[!htb]
\includegraphics[width=1.0\columnwidth]{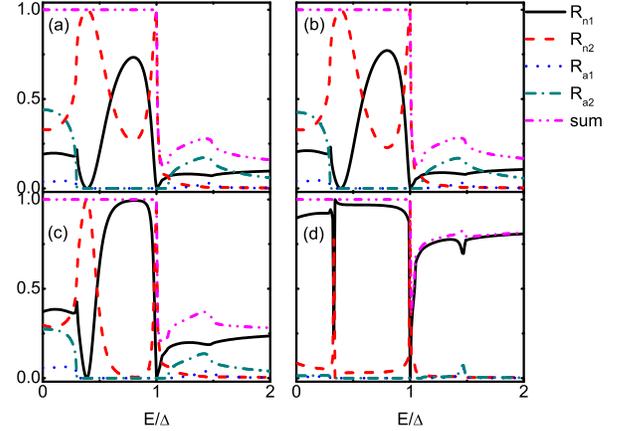}
\caption{The incident energy $E$ dependence of reflection probabilities for $E_{z}=0.4\Delta$, $\mu_{N}=\mu_{S}=0.7\Delta$ and $V_{0}=10$ with (a) $\theta=0$, (b) $\theta=0.1\pi$, (c) $\theta=0.3\pi$, and $\theta=0.45\pi$.}
\label{fig6}
\end{figure}

Now, we turn to the incident energy dependence of reflection probabilities.
Fig.\ref{fig6} presents the results for different $\theta$ at $\mu_{N}=\mu_{S}=0.7\Delta$ and $E_{z}=0.4\Delta$.
The gap for HLQs spans from $0.3\Delta$ to $1.1\Delta$.
As a result, the Andreev reflections, both RAR and SAR,
are absent in the energy range $0.3\Delta<E<1.1\Delta$.
The double Andreev reflections can happen in the other energy range,
$E<0.3\Delta$ or $E>1.1\Delta$, as shown in Figs.\ref{fig6} (a)-(d).
For $E<0.3\Delta$, the double Andreev reflections originate
from the involved conduction band $E_{h}^{-}$.
On the other hand, while for $E>1.1\Delta$,
the double Andreev reflections originate from
the involved valence band $E_{h}^{+}$.
For $E<0.3\Delta$, the changes of RAR and SAR are not dramatic.
When the incident energy $E$ increases from $1.1\Delta$,
the curves of RAR and SAR rise first and then gradually decrease.
Note, the summation of the reflection probabilities for $E>\Delta$
will not equal to $1$ due to the emergence of the quasiparticle transmissions.
As for the normal reflections, SNR $R_{n1}$ and RNR $R_{n2}$ oscillate
dramatically in the range $0.3\Delta<E<\Delta$
while the changes in the range $E<0.3\Delta$ and $E>1.1\Delta$ are not obvious.
In addition, the Andreev reflections, both RAR and SAR,
will tend to zero as $\theta$ approaches to $0.5\pi$
as shown in Fig.\ref{fig6} (d).
This is consistent with the incident angle dependence of
reflections discussed above.

\begin{figure}[!htb]
\includegraphics[width=1.0\columnwidth]{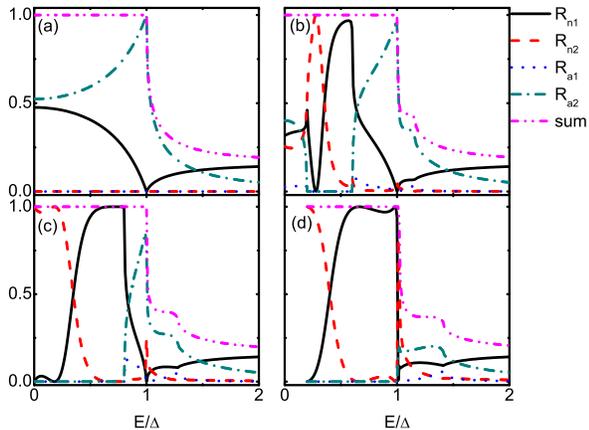}
\caption{
The incident energy $E$ dependence of the reflection probabilities
for $\theta=0.2\pi$, $\mu_{N}=\mu_{S}=0.4\Delta$ and $V_{0}=10$ with
(a) $E_{z}=0$, (b) $E_{z}=0.2\Delta$, (c) $E_{z}=0.4\Delta$, and (d) $E_{z}=0.6\Delta$.}
\label{fig7}
\end{figure}

Fig.\ref{fig7} presents the incident energy dependence of reflections
for different $E_{z}$ at $\mu_{N}=\mu_{S}=0.4\Delta$
with the fixed incident angle $\theta=0.2\pi$.
For $E_{z}=0$, there are only SNR and RAR in the scattering processes
due to the orbital coupling is absent (see Fig.\ref{fig7}(a)).
For $E_{z}=0.2\Delta$, the gap for HLQs is from $0.2\Delta$ to $0.6\Delta$.
Correspondingly, the Andreev reflections are absent in the range $0.2\Delta<E<0.6\Delta$ (see Fig.\ref{fig7}(b)).
In this energy range, the behaviours of the normal reflections
are similar to the results in Fig.\ref{fig6}.
In the energy range $0.6\Delta<E<\Delta$,
the probability of SNR drops sharply and
RAR $R_{a2}$ dominates the scattering processes.
For $E_{z}=0.4\Delta$, the gap for HLQs is from $0$ to $0.8\Delta$
and there are only normal reflections in the energy range $E<0.8\Delta$
(see Fig.\ref{fig7} (c)).
For $E_{z}=0.6\Delta$, the gap for HLQs spans from $0$ to $\Delta$,
the Andreev reflections will be absent
inside the whole gap (see Fig.\ref{fig7} (d)).

\begin{figure}[!htb]
\includegraphics[width=1.0\columnwidth]{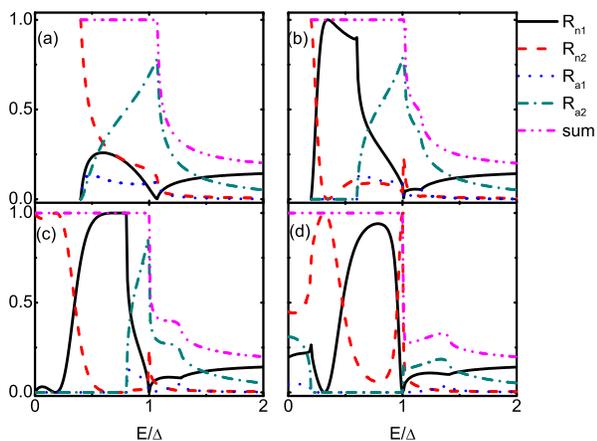}
\caption{The incident energy $E$ dependence of the reflection probabilities
for $\theta=0.2\pi$, $E_{z}=0.4\Delta$ and $V_{0}=10$ with
(a) $\mu_{N}=\mu_{S}=0$, (b) $\mu_{N}=\mu_{S}=0.2\Delta$,
(c) $\mu_{N}=\mu_{S}=0.4\Delta$, and (d) $\mu_{N}=\mu_{S}=0.6\Delta$.}
\label{fig8}
\end{figure}

Fig.\ref{fig8} gives the reflection probabilities for different $\mu_{N}$ and $\mu_{S}$ at $E_{z}=0.4\Delta$ and $\theta=0.2\pi$. For $\mu_{N}=\mu_{S}=0$, the gaps for ELQs and HLQs are both from $-0.4\Delta$ to $0.4\Delta$ which are symmetric about $E=0$. For $E<0.4\Delta$, the energy of the incident quasiparticles lies in the gaps and there is no traveling wave for ELQs. The reflections, both the normal ones and Andreev ones, are not present as shown in Fig.\ref{fig8} (a). When $E>0.4\Delta$, ELQs on the conduction band $E_{e}^{+}$ and HLQs on the valence band $E_{h}^{+}$ participate in the scattering processes. Double normal reflections and double Andreev reflections exist simultaneously. It is worth noting that the summation of reflection probabilities equals to $1$ from $E=0.4\Delta$ to $E\approx 1.077\Delta$ not to $\Delta$ (see Fig.\ref{fig8}(a)). This indicates no quasiparticle transmissions in the energy range $\Delta<E<1.077\Delta$ and the superconducting gap in SC is larger than $\Delta$ in this case. Actually, the quasiparticle dispersion in SC is $E=\sqrt{\epsilon_{k}^2+E_{z}^2+\Delta^2}$ when $\mu_{S}=0$. The effective gap in SC is $\sqrt{E_{z}^2+\Delta^2}$ not $\Delta$ which is consistent with the results in Fig.\ref{fig8}(a).

When $\mu_{S}\ne0$, the quasiparticle dispersion in SC becomes $E=\sqrt{[\sqrt{\epsilon_{k}^2+E_{z}^2}-\mu_{S}]^2+\Delta^2}$. When $\mu_{S}<E_{z}$, the effective gap in SC is $\sqrt{[\vert E_{z}\vert-\mu_{S}]^2+\Delta^2}$.
For example, if $E_{z}=0.4\Delta$ and $\mu_{S}=0.2\Delta$
as considered in Fig.\ref{fig8}(b),
the effective gap will approximate to $1.02\Delta$ (see also Fig.\ref{fig7}(d)).
At the same time, there will be a gap from $-0.6\Delta$
to $0.2\Delta$ for ELQs in NLSM.
Hence, there are no reflections in the range $0<E<0.2\Delta$ and
the sum of probabilities for four reflections equals to $1$ in the range $0.2\Delta<E<
\sqrt{[\vert E_{z}\vert-\mu_{S}]^2+\Delta^2} \simeq 1.02\Delta$.
When $\mu_{S}$ is raised to $\mu_{S}=E_{z}$,
the gap in SC reverts to its intrinsic value $\Delta$
and the gap for ELQs in NLSM completely locates below $E=0$,
which leads to the reflections probabilities shown in Fig.\ref{fig8}(c).
Here the Andreev reflections disappear at $E<0.8\Delta$
due to the gap for HLQs in NLSM.
When $\mu_{S}$ continues to rise,
the gap in SC will not be changed and keep the value $\Delta$.
For $\mu_{N}=\mu_{S}=0.6\Delta$ in Fig.\ref{fig8}(d),
ELQs and HLQs involved in the scattering are from the conduction
bands $E_{e}^{+}$ and $E_{h}^{-}$ in the energy range $0<E<0.2\Delta$
while for $0.2\Delta<E<\Delta$, only ELQs from $E_{e}^{+}$
participate in the scattering.

\begin{figure}[!htb]
\includegraphics[width=1.0\columnwidth]{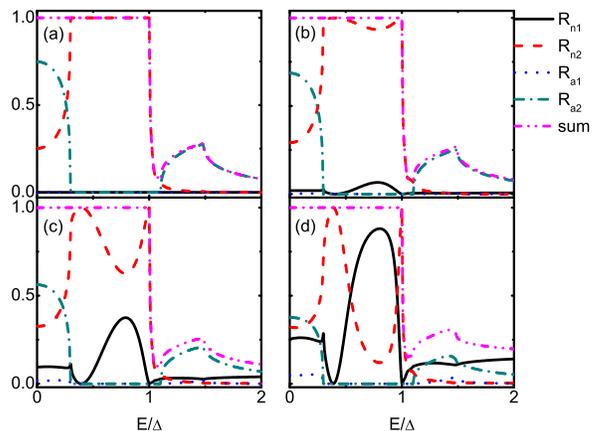}
\caption{The incident energy $E$ dependence of the reflection probabilities
for $\theta=0.2\pi$, $\mu_{N}=\mu_{S}=0.7\Delta$ and $E_{z}=0.4\Delta$
with (a) $V_{0}=0$, (b) $V_{0}=2$, (c) $V_{0}=5$, and $V_{0}=10$.}
\label{fig9}
\end{figure}

Fig.\ref{fig9} shows the incident energy dependence of reflections
for different interfacial barriers
at $\mu_{N}=\mu_{S}=0.7\Delta$ and $E_{z}=0.4\Delta$.
Since the relation $\mu_{N}>E_{z}$ is fixed,
the conduction band $E_{e}^{+}$ is always involved
in the scattering processes while the dispersions for HLQs have
a gap from $0.3\Delta$ to $1.1\Delta$.
For $0.3\Delta<E<1.1\Delta$, there will be no Andreev reflections
as given in Fig.\ref{fig9} (a)-(d).
When the interface is transparent, the specular reflections,
SNR and SAR, are also absent.
Therefore, there will be only RNR in the energy range $0.3\Delta<E<1.1\Delta$
for $V_{0}=0$ with $R_{n2}=1$, as shown in Fig.\ref{fig9}(a).
In this case, the incident ELQs are completely reflected back
along the retro-reflected direction, which phenomenon is very rare.
In the energy range $E<0.3\Delta$, the conduction band $E_{h}^{-}$
also participates in the scattering and two retro-reflections coexist.
This is consistent with the results in Fig.\ref{fig4}(a).
When $V_{0}\ne0$, the specular reflections are activated.
For $0.3\Delta<E<1.1\Delta$, there will be double normal reflections
and for $E<0.3\Delta$ or $E>1.1\Delta$, there will be four types of reflections.
From Figs.\ref{fig9} (b)-(d), we can find the specular reflections
are enhanced as the opacity of the interface is increased.
In other words, the presence of finite barrier height is beneficial
to the specular reflections and detrimental to the retro-reflections.
This conclusion is consistent with the results in Figs.\ref{fig4} (b)-(d).

\subsubsection{\label{subsubsec2.2.3}Conductance spectra}

\begin{figure}[!htb]
\includegraphics[width=1.0\columnwidth]{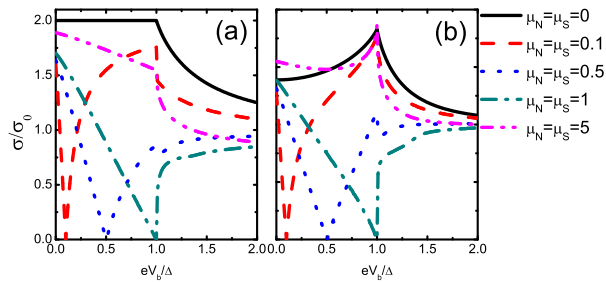}
\caption{The normalized conductance spectra for different
chemical potentials with (a) the barrier potential $V_{0}=0$ and (b) $V_{0}=10$.}
\label{fig10}
\end{figure}

Here, we discuss the conductance of the NLSM-SC junctions.
In the expression of conductance in Eq.(\ref{cond}),
there is an integral of $E_{z}$ from $-eV_b$ to $e V_b$ with $V_b$ the bias.
Since the probabilities are even functions of $E_{z}$,
it is enough to analyze the conductance supposing $E_{z}>0$.
Fig.\ref{fig10}(a) gives the normalized conductance
for the transparent interface with the barrier potential $V_{0}=0$.
According to the discussions in the last subsection,
the specular reflections in this situation are absent
and only RNR and RAR exist.
For $\mu_{N}=0$, the gaps for ELQs and HLQs in NLSM are symmetric
about the energy $E=0$ and span from $-E_{z}$ to $E_{z}$.
The expression of conductance requires $eV_b>E_{z}$,
which indicates the conduction band $E_{e}^{+}$
and the valence band $E_{h}^{+}$ always participate in the scattering processes.
Inside the gap, the total RAR will lead to the normalized conductance
being $2$. This is the same value with the conductance
of the ordinary normal-metal-SC junctions with the transparent interface \cite{Blonder}.

For $\mu_{N}=\mu_{S}\ne0$, HLQs are not always involved in the scattering.
The gap for HLQs is no longer symmetric about $E=0$
and spans from $\mu_{N}-E_{z}$ to $\mu_{N}+E_{z}$.
The condition for RAR is $eV_b+E_{z}<\mu_{N}$ or $eV_b-E_{z}>\mu_{N}$
(Note, SAR is absent due to $V_{0}=0$.).
In the former situation, HLQs from the conduction
band $E_{h}^{-}$ are responsible for RAR.
For $eV_b<\mu_{N}$, when $eV_b$ is increased,
the value range of $E_{z}$ satisfying the former condition becomes narrow.
Correspondingly, the $E_{z}$ range for RAR also narrows
and the conductance decreases.
In the latter situation, HLQs from the valence band $E_{h}^{+}$
are responsible for RAR.
For $eV_b>\mu_{N}$, when $eV_b$ is increased,
the value range of $E_{z}$ satisfying the latter condition becomes wide.
The $E_{z}$ range for RAR also widens and the conductance rises.
For the special case of $eV_b=\mu_{N}$,
no value of $E_{z}$ under the supposed relation $E_{z}>0$
meets the two conditions for RAR.
There is only the total RNR in the scattering processes
which will cause the zero conductance.
Our analysis can be immediately demonstrated by the numerical results with $\mu_{N}=\mu_{S}=0.1\Delta$, $0.5\Delta$ and $1.0\Delta$ in Fig.\ref{fig10}(a).
For $\mu_{N}=\mu_{S}> \Delta$, the zero conductance phenomenon
will not disappear, since quasiparticle transmissions happen
at the incident energy larger than $\Delta$.

Fig.\ref{fig10}(b) gives the normalized conductance for $V_{0}=10$.
According to the discussions in the last subsection,
the specular reflections will be activated.
For $\mu_{N}=0$, double normal reflections and
double Andreev reflections coexist.
The presence of the normal reflections will weaken the conductance
and its value can not reach $2\sigma_{0}$ inside the gap.
For $\mu_{N}\ne0$, the condition for the Andreev reflections,
both RAR and SAR, is also $eV_b+E_{z}<\mu_{N}$ or $eV_b-E_{z}>\mu_{N}$.
For $eV_b=\mu_{N}$, the conductance becomes zero.
On the both sides of $eV_b=\mu_{N}$, the variations of conductance
are similar to the curves in Fig.\ref{fig10}(a).
The conductance for $\mu_{N}=\mu_{S}=5\Delta$ at $V_{0}=10$
is also presented in Fig.\ref{fig10}(b) and its properties are
similar to the curves with $\mu_{N}=\mu_{S}=0$.
Although the impressive vanishing conductance at $eV_{b}=\mu_{N}$
are reminiscent of graphene-SC junctions \cite{Beenakker},
the chemical potential dependence of conductance spectra is very different.
For the case of graphene, the single Andreev reflection, RAR or SAR,
happens depending on the value of the chemical potential.

\section{\label{sec3}Nodal-line parallel to interface}
\subsection{\label{subsec3.1} Model and formalism}
\begin{figure}[!htb]
\includegraphics[width=1.0\columnwidth]{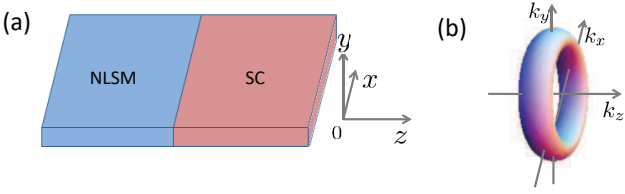}
\caption{
(a) Schematic illustration of the NLSM-SC junction with the nodal-line parallel to the interface. The interface is located at $z=0$ and parallel to the $xy$ plane. (b) The torus-like isoenergetic surface in NLSM. The crystal axis $k_{z}$ is perpendicular to the interface of the junction.}
\label{ap1}
\end{figure}

In this subsection, we give the formalism for the scattering processes and conductance of the NLSM-SC junction
with the nodal-line parallel to the interface as shown in Fig.\ref{ap1}.
In this case, the NLSM and SC are located in the regions $z<0$ and $z>0$, respectively, with the NLSM-SC interface being at $z=0$.
The transport is along the $z$ direction.
The Hamiltonians of the NLSM and SC are the same as the Eqs.(2) and (8)
in Sec.\ref{subsec2.1}.
The quasiparticle dispersions of the NLSM for ELQs and HLQs solved
from the BdG equation are
\begin{equation}
E_{e}^{\pm}=\pm\sqrt{\epsilon_{k}^2+\hbar^2v^2k_{z}^2}-\mu_{N},
\end{equation}
and
\begin{equation}
E_{h}^{\pm}=\pm\sqrt{\epsilon_{k}^2+\hbar^2v^2k_{z}^2}+\mu_{N},
\end{equation}
respectively, which are the same as the dispersions in Eqs.(3) and (4).
However, for each band, there is only one type of quasiparticle.
Taking the conduction band $E_{e}^{+}$ as an example,
when the incident energy $E$ and wave vectors $k_x$ and $k_y$
are given, the equation $E_{e}^{+}(k_x, k_y, k_z) =E$ of $k_z$
has only two solutions, $k_z^{e}$ and $-k_z^{e}$.
Here the $k_z^{e}$ and $-k_z^{e}$ states propagate along the $+z$
and $-z$ directions, respectively.
This is different from the case of
the nodal-line perpendicular to the interface, in which
there are two types of quasiparticle for each band.
So for the present case, there are only two reflection processes,
the normal reflection and Andreev reflection,
when a beam of ELQs are injected from NLSM side.
Consider that the incident ELQs are from the conduction band $E_{e}^{+}$.
They will be specularly reflected as ELQs on the band $E_{e}^{+}$
and retro-Andreev reflected as HLQs on the conduction band $E_{h}^{-}$
if $E<\mu_{N}$ or
specular Andreev reflected as HLQs on the valence band $E_{h}^{+}$ if $E>\mu_{N}$.
In this situation, the scattering processes are the same
as that in the graphene-SC junction.\cite{Beenakker,addsun3,addsun4}

Assuming an ELQ with the wave vector $k_{z}^{e}$ is injected from NLSM, the wave function in NLSM can be written as
\begin{equation}
\begin{split}
\Psi_{N}&=\left(
\begin{array}{cc}
i\xi_{1}\\
1\\
0\\
0\\
\end{array}
\right)e^{ik_{z}^{e}z}
+
r_{n}\left(
\begin{array}{cc}
-i\xi_{1}\\
1\\
0\\
0\\
\end{array}
\right)e^{-ik_{z}^{e}z}\\
&+
r_{a}\left(
\begin{array}{cc}
0\\
0\\
-\alpha i\xi_{2}\\
1
\end{array}
\right)e^{\alpha ik_{z}^{h}z},
\end{split}
\end{equation}
where $\xi_{1}=\frac{E+\mu_{N}+\Omega_{N}^{+}}{\hbar vk_{z}^{e}}$, $\xi_{2}=\frac{E-\mu_{N}-\Omega_{N}^{-}}{\hbar vk_{z}^{h}}$ and $k_{z}^{e(h)}=\sqrt{\frac{2m}{\hbar^2}\{\sqrt{(e_{k}+mv^{2})^2+[(E+(-)\mu_{N})^2-e_{k}^{2}]}-(e_{k}+mv^2)\}}$ with $\Omega_{N}^{+(-)}=\sqrt{(E+(-)\mu_{N})^2-\left(\hbar v k_{z}^{e(h)}\right)^2}$ and $e_{k}=\frac{\hbar^2}{2m}(k_{x}^2+k_{y}^2)-E_{0}$. When $E<\mu_{N}$, $\alpha=1$ and $r_{a}$ denotes the RAR amplitude; when $E>\mu_{N}$, $\alpha=-1$ and $r_{a}$ denotes the SAR amplitude.

The wave function in SC is
\begin{equation}
\begin{split}
\Psi_{S}&=c\left(
\begin{array}{cc}
iu\zeta_{1}\\
u\\
i\zeta_{1}\\
1\\
\end{array}
\right)e^{ik_{z}^{+}z}
+
d\left(
\begin{array}{cc}
-iv\zeta_{2}\\
v\\
-i\zeta_{2}\\
1\\
\end{array}
\right)e^{-ik_{z}^{-}z},
\end{split}
\end{equation}
where $u=\frac{E+\Omega_S}{\Delta}$, $v=\frac{E-\Omega_S}{\Delta}$, $\zeta_{1}=\frac{\mu_{S}+\Omega_S+\Omega_{S}^{+}}{\hbar vk_{z}^{+}}$,
$\zeta_{2}=\frac{\mu_{S}-\Omega_S+\Omega_{S}^{-}}{\hbar vk_{z}^{-}}$ and $k_{z}^{\pm}=\sqrt{\frac{2m}{\hbar^2}\{\sqrt{(e_{k}+mv^2)^2+(\mu_{S}\pm\Omega_S)^2-e_{k}^2}-(e_{k}+mv^2)\}}$ with $\Omega_S=\sqrt{E^2-\Delta^2}$ and $\Omega_{S}^{\pm}=\sqrt{(\mu_{S}\pm\Omega_S)^2-\hbar^2 v^2{k_{z}^{\pm}}^2}$.

We are concerned with the nearly linear regime of the dispersion $E(k_{z})$
where the parameter $mv^2$ is large enough.
In this situation, the following boundary condition
for the wave functions are suitable,
\begin{equation}
\psi_{N}(z=0^{-})=\psi_{S}(z=0^{+}).
\end{equation}
The reflection amplitudes $r$ and $r_{a}$ can be obtained
from the boundary condition and the reflection probabilities are given by
\begin{equation}
\begin{split}
R_{n}(E)&=\vert r\vert^2,\\
R_{a}(E)&=\big| \frac{\xi_{2}}{\xi_{1}} \big|\vert r_{a}\vert^2.
\end{split}
\end{equation}
Inside the gap with $E<\Delta$, the conservation relation $R_{n}+R_{a}=1$ is satisfied.

Since the traveling wave solution for ELQs is restricted in the interval $-(eV_b+\mu_{N})<e_{k}< eV_b+\mu_{N}$, the conductance can be expressed as
\begin{equation}
\sigma=\frac{2e^2}{h}\frac{S}{2\pi}\int_{k_{min}}^{k_{max}}
[1+R_{a}(eV_b)-R_{n}(eV_b)]kdk,
\end{equation}
with $k_{min}=\sqrt{\frac{2m}{\hbar^2}[E_{0}-(eV_b+\mu_{N})]}$ and $k_{max}=\sqrt{\frac{2m}{\hbar^2}[E_{0}+(eV_b+\mu_{N})]}$, $V_b$ is the bias
between the NLSM and SC.
The normalized conductance is $\sigma/\sigma_{0}$ with $\sigma_{0}$ the conductance of the NLSM-NLSM junction.
\subsection{\label{subsec2.2}Results and discussions}
\begin{figure}[!htb]
\includegraphics[width=1.0\columnwidth]{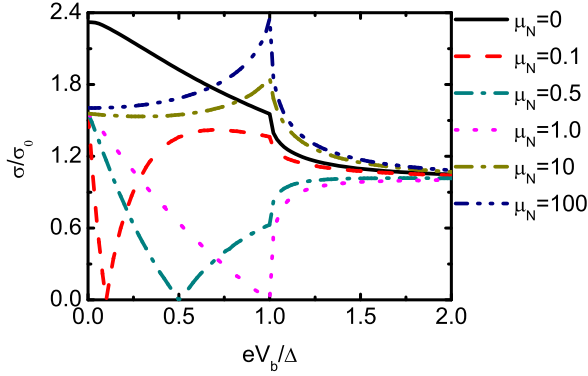}
\caption{
The normalized conductance spectra for the junction with
the nodal-line parallel to the interface.
The parameters are taken as $E_{0}=200\Delta$ and $\mu_{S}=500\Delta$.}
\label{fig11}
\end{figure}

We present the conductance of the NLSM-SC junction
with the nodal-line parallel to the NLSM-SC interface
for various of $\mu_{N}$ in Fig.\ref{fig11}.
Now, the $k_{z}$ axis is perpendicular to the interface
and the current flows along the $z$ axis.
The nearly linear dispersion regime of $E$ about $k_{z}$ is concerned,
when the parameter $mv^2$ is large enough.
In this situation, there are only SNR and a single Andreev reflection,
RAR or SAR, in the scattering processes.
For the Andreev reflection, RAR happens for $eV_b<\mu_{N}$
while SAR happens for $eV_b>\mu_{N}$.
Inside the gap, RAR dominates the conductance if $\mu_{N}\gg\Delta$,
while SAR dominates the conductance if $\mu_{N}\ll\Delta$.
The scattering processes are the same with those in the graphene-SC junction\cite{Beenakker}.
Actually, the conductance spectra in Fig.\ref{fig11} are highly similar
to the results for the graphene junction as shown in Ref.[\onlinecite{Beenakker}].
From the results in Figs.\ref{fig10} and \ref{fig11},
it is found that the transport properties of the NLSM-SC junction
exhibit anisotropy. The conductance spectra depend on the orientation
relationship between the crystal axis (or the nodal-line)
and the junction interface.

\section{\label{sec4}Conclusions}
We study the scattering processes and conductance in NLSM-SC junctions.
We find the novel quadruple reflections,
double normal reflections and double Andreev reflections,
can happen simultaneously in NLSM.
The probabilities of reflections are systematically
studied under different parameters.
The occurrence and disappearance of the reflections and their magnitudes
can be controlled by changing the incident quasiparticle energy,
the incident angle, the interfacial barrier, the chemical potentials
and the orbital coupling strength.
Distinctive features manifest themselves in the conductance spectra
and the spectra show anisotropy depending on the mutual
orientation of the nodal-line and the interface.
This unique quadruple reflections may be useful for the engineering
of the electronic beam splitter based on the ballistic system.

\section*{\label{sec5}ACKNOWLEDGMENTS}
This work was financially supported by National Key R and D Program of China (2017YFA0303301),
NSF-China under Grants Nos. 11921005 and 11447175,
the Strategic Priority Research Program of Chinese Academy of Sciences (XDB28000000)
and the Natural Science Foundation of Shandong Province under Grants No. ZR2017QA009.

\end{document}